\documentclass[useAMS,usenatbib,twocolumn]{mn2e}

\usepackage{dcolumn}
\usepackage{graphicx}
\usepackage{url}
\usepackage{color}
\usepackage{epsfig}

\newcommand{\cm}{cm$^{-1}$}

\newcommand{\eqref}[1]{(\ref{#1})}

\newcommand{\angmol}{{\sc angmol}}

\title[ExoMol: XI The spectrum of nitric acid]{ExoMol molecular line lists: XI The spectrum of nitric acid}

\date{\today}
\author {A.I. Pavlyuchko$^{a,b}$, S.N. Yurchenko$^a$, Jonathan Tennyson$^a$ \\
$^a$ Department of Physics and Astronomy, University College London, London, WC1E 6BT, UK;\\
$^b$ Moscow State University of Civil Engineering (MGSU), Russia, (pavlyuchko@rambler.ru)}

\date{Accepted XXXX. Received XXXX; in original form XXXX}

\pagerange{\pageref{firstpage}--\pageref{lastpage}} \pubyear{2012}
\begin{document}
\maketitle
\begin{abstract}
  Nitric acid is a possible biomarker in the atmospheres of
  exoplanets.  An accurate line list of rotational and rotational-vibrational
transitions is computed for nitric acid
  (HNO$_3$).  This line list covers wavelengths longer than 1.42 $\mu$m
  (0 -- 7000 cm$^{-1}$) and temperatures up to 500 K. The line list is computed using a
  hybrid variational -- perturbation theory and empirically tuned
  potential energy and dipole surfaces.  It comprises almost 7 billion transitions
  involving rotations up to $J=100$. Comparisons with spectra from the
  HITRAN and PNNL databases demonstrate the accuracy of our
  calculations. Synthetic spectra of water -- nitric acid mixtures
  suggest that nitric acid has features at 7.5 and 11.25 $\mu$m that are
  capable of providing a clear signature for HNO$_3$; the feature
at 11.25 $\mu$m is particularly promising. Partition
  functions plus full line lists of transitions are made available in
  an electronic form as supplementary data to the article and at
  \url{www.exomol.com}.

\end{abstract}

\label{firstpage}

\section{Introduction}

The infrared spectrum of the nitric acid molecule, HNO$_3$, is of
astrophysical interest because of the growing interest in the study
the of atmospheres of extrasolar planets. In particular, this interest
is particularly linked to that of the planets with earth-like
atmospheres with a high content of nitrogen and oxygen. Nitric acid
can be clearly observed in the earth's atmosphere from space
\citep{11CoMaSa.HNO3} as it has a number of features which lie in
water transparency windows. Its spectrum constitutes a possible
biomarker as its existence is indicative of the presence of free oxygen and
nitrogen, both of which lack strong spectral signatures at long
wavelengths. It is also thought likely to pay an important role in
nitrogen fixation on Mars and other earth-like planets
\citep{07SuKh.HNO3}. It has also been suggested that $\gamma$-ray
bursts could produce large quantities of HNO$_3$ in the atmospheres of
earth-like planets \citep{06ThMe.HNO3}.  Nitric acid is a constituent
of atmospheric ices on Earth and is thought likely to be present in
the ice crust of Europa \citep{11CoMaSa.HNO3}.  Its formation in
interstellar ammonia ices has also been suggested
\citep{13ZaRoGa.HNO3}, although nitric acid has yet to be observed in
the interstellar medium.

Given the importance of HNO$_3$ in the earth's atmosphere, its infrared
spectrum has been well-studied in the laboratory. However, many of
these spectra remain incompletely analysed with very few assignments
to transitions at wavelengths shorter than 5 $\mu$m. This lack of
assignment means that the spectra cannot be used for spectral
simulations at temperatures other than the one of the original
experiment.  Spectral data on HNO$_3$ is only given in the HITRAN
database \citep{jt557} for up to 1770 \cm\ (wavelengths longer than 5.6
$\mu$m). At shorter wavelength the Pacific Northwest National
Laboratory (PNNL) database \citep{PNNL} provides infrared cross
sections, but again these are only valid at the temperature for which
they are recorded. Our results compare favourably with these
sources, although this comparison shows both have their limitations \citep{jt603}.

The ExoMol project aims at generating comprehensive line observation
and modelling atmospheres of exoplanets and other hot astronomical
objects such as brown dwarfs and cool stars; its aims, scope and
methodology have been summarised by \citet{jt528}. The project has
provided rotation-vibration line lists for a number of polyatomic
molecules such as HCN \citep{jt570}, H$_2$S \citep{jth2s}, PH$_3$
\citep{jt592}, H$_2$CO \citep{jt597} and CH$_4$ \citep{jt564,jt572}.
None of these molecules contains more than two heavy atoms and the
only pentatomic molecule considered, methane, has four hydrogens and
high symmetry. It is clear that computing a comprehensive,
temperature-dependent line list for HNO$_3$ represents a considerable
computational challenge. To address this challenge we have developed a
hybrid variational-perturbation theory procedure for computing spectra
of such molecules \citep{jt588} and have tested this for
room-temperature nitric acid spectra \citep{jt603}. In this work we
present a comprehensive line list for nitric acid which should be
valid for temperatures up to 500 K. Given that nitric acid is only
likely to exist in atmospheres which also contain water, we explicitly
consider regions where HNO$_3$ spectra are likely to be observable in a
humid atmosphere.

\section{Method}

Rotation-vibration line lists were generated using the program
\angmol\ \citep{88GrPa.method,jt588}.  Although \angmol\ is designed
to solve the nuclear motion problem for polyatomic molecules using a
hybrid variational-perturbation theory method, it actually provides a
complete environment for performing the line list calculations.
\angmol\ automatically generates the inputs required to drive the
appropriate {\it ab initio} electronic structure program, here the quantum chemistry
package {\sc MOLPRO} \citep{12WeKnKn.methods}, to provide the necessary
inputs for the nuclear motion calculations (potential energy, Hessian matrix
and dipole moments). It also automatically adjusts both the potential
energy surface (PES) and dipole moment function (DMF), which are represented as Taylor-expansions
about equilibrium, to reproduce observed line positions and intensities
using the method of regularisation \citep{79TiArxx.method, 92GrDexx.method}.
This method
uses constraints provided by the initial {\it ab initio} calculations to allow many
more parameters to be varied than there are experimental data. For full
details of the fits and associated surfaces see \citet{jt603}.

\angmol\ solves a Watson-like nuclear motion Hamiltonian expressed in
internal curvilinear vibrational coordinates in three steps.  First
the rotation-less ($J=0$) vibrational problem is solved in a basis of
Morse oscillators for the stretches and harmonic oscillators for the
bends by direct diagonalisation of a Hamiltonian matrix for low-lying
vibrational states with contributions from higher states (basis
functions) included using perturbation theory.  In the second step,
rotational problems for each vibrational state are solved using the
eigenvectors from the vibrational problem and Wigner rotation
matrices. An Eckart embedding is used to minimise Coriolis
interactions which are again introduced using perturbation theory. The
final step computes transition intensities; this step reduces the size
of the calculation by predicting which transitions will be too weak to
make a significant contribution and not calculating them.  Transitions
whose intensity is less than $10^{-12}$ km/mole ($\approx  10^{-31}$ cm/molecule.) were neglected.
Integrals
over the kinetic energy operator, PES and DMF are all simplified by
the use of Taylor expansions for the operators. Here the PES and DMF
were represented as fourth-order and second-order expansions,
respectively. Further details of the calculation can be found in
\citep{jt603}.

\begin{figure}
\begin{center}
{\leavevmode \epsfxsize=7.0cm \epsfbox{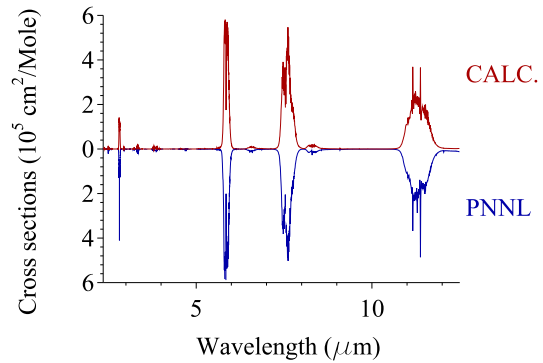}}
\caption{Comparison of our computed spectrum (Calc) with cross sections from the PNNL
database \citep{PNNL} at a temperature of 298 K. Cross sections were generated
from our line list using a Voigt profile with a Doppler and pressure broadening widths
both set to 0.075 cm$^{-1}$.}
\label{fig:PNNL}
\end{center}
\end{figure}

Our aim was to compute a HNO$_3$ line list which is complete up to
7000 cm$^{-1}$ (longwards of 1.42 $\mu$m) for temperatures up to 500
K. For this we considered energy levels up to 9000 cm$^{-1}$ and
rotational states with $J$ up to 100. $J=100$ corresponds to a
rotational excitation of about 2130 cm$^{-1}$.

The initial vibrational ($J=0$) problem was solved using vibrational
basis sets with polyad number 14. By this we mean that all
combinations of functions were included when the sum of the orders of
the polynomials was less than or equal to 14. The size of the
basis required increases combinatorially with polyad number $N$:
for HNO$_3$, which has 9 vibrational modes, the size of the basis
is given by the binomial coefficient $_{N+9}C_9$ which equals 817~190
for $N=14$ and 48~620 for $N=9$.
Matrices constructed
with polyad 9 were explicitly diagonalised once the effects of higher
states were included using perturbation theory, see \citet{jt603} for
further details. These calculations gave a total of 22~049 vibrational
states below 9000 cm$^{-1}$.

Rotationally excited states were computed  based on calculation
which  only considered up to polyad 7 but
with the vibrational band origins fixed by the larger $J=0$
calculation. This gave 9477 states lying below 7000 cm$^{-1}$ of which
intensity considerations (see \citet{jt603} for details) showed that
only 1715 need to be considered explicitly. No transition whose
intensity was less than $10^{-31}$ cm/molecule was considered.

Our final line list links 17~494~715 vibration-rotation energy levels
with  6~722~136~109 transitions.


Temperature-dependent partition functions for HNO$_3$ were calculated
by explicitly summing all the calculated energy levels.
Table~\ref{tab:pf} compares our results with those used by HITRAN
\citep{03FiGaGo.partfunc}.  The agreement is very good; our results
are systematically slightly larger by less than 0.2\%\ at low
temperatures rising to 0.9 \%\ at 500 K. This increase is probably due
to our better treatment of anharmonic effects, particularly in the
large-amplitude, low-frequency OH torsional mode. A file containing the
partition function 1 K steps for temperatures up to 500 K  is given in
supplementary material.

\begin{table}
\caption{Partition function, $Q(T)$, HNO$_3$, as a function of temperature.}
\label{tab:pf}
\begin{center}
\begin{tabular}{lrrrr}
\hline\hline
T   &	Q(T) & Q(T)    \\
 K   &  This work & \citet{03FiGaGo.partfunc}   \\
\hline
   60 &  15032.00 &15010\\
  110 &  37418.02 &37374   \\
  160 &  67172.12 &67105 \\
  210 & 107006.55 &106880  \\
  260 & 161987.99 &161650 \\
  310 & 239197.27 &238250  \\
  360 & 348081.08 &345830  \\
  410 & 500933.14 &496570    \\
  460 & 713230.66 &706730\\
\hline
\end{tabular}
\end{center}
\end{table}

\section{Results}

The line list contain almost  7 billion transitions.
For compactness and ease of use, it is  divided into separate energy level and
transitions file. This is done using standard ExoMol format \citep{jt548} which
is based on a method originally developed for the BT2 line list \citep{jt378}.
Extracts from the start of the files are given in
Tables~\ref{tab:levels} and \ref{tab:trans}.  The full line list  can be downloaded from the CDS, via
\url{ftp://cdsarc.u-strasbg.fr/pub/cats/J/MNRAS/xxx/yy}, or
\url{http://cdsarc.u-strasbg.fr/viz-bin/qcat?J/MNRAS//xxx/yy}.  The line lists
and partition function, as well as the absorption spectrum
given in cross section format \citep{jt542}, can all be obtained from there as
well as at \url{www.exomol.com}.

\begin{figure}
\begin{center}
{\leavevmode \epsfxsize=7.0cm \epsfbox{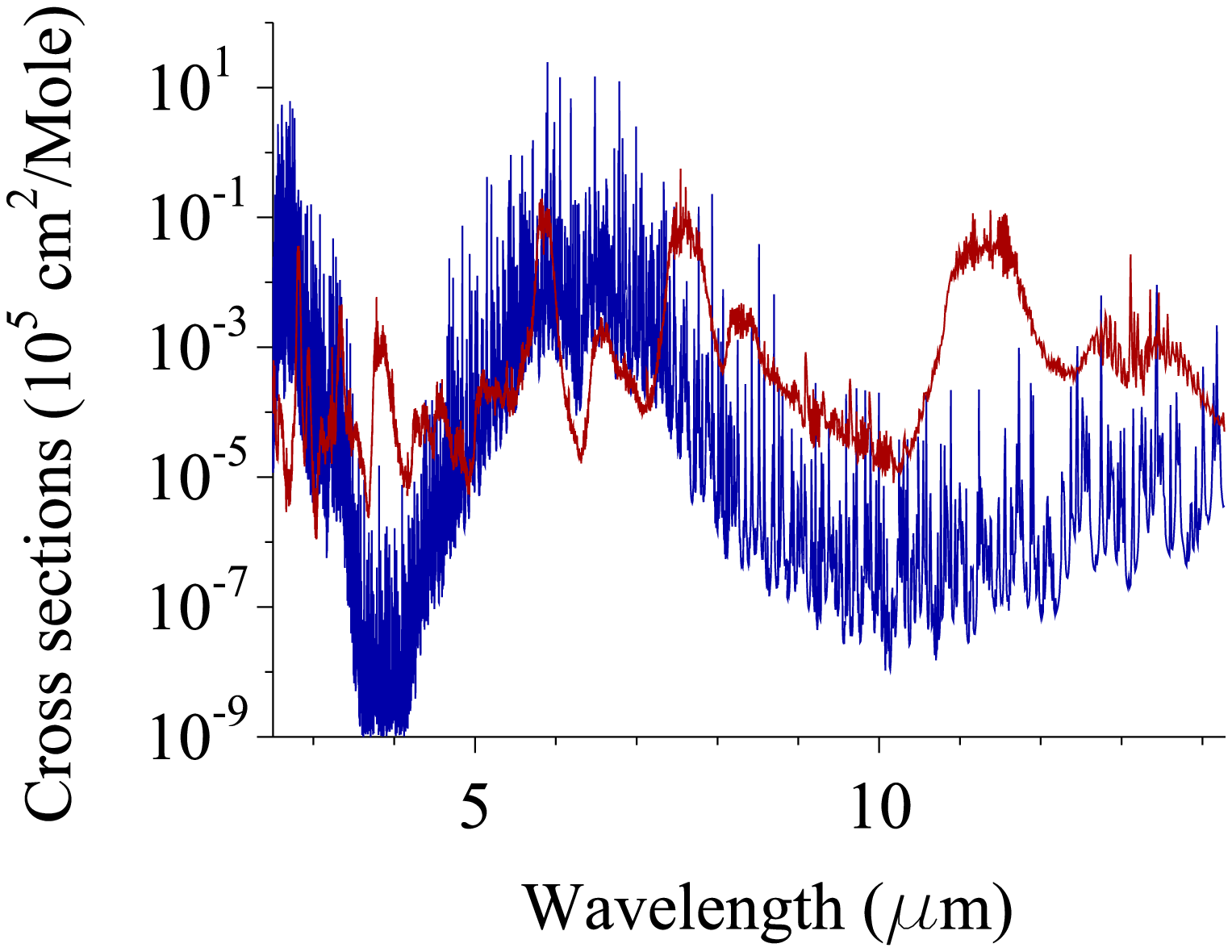}}
{\leavevmode \epsfxsize=7.0cm \epsfbox{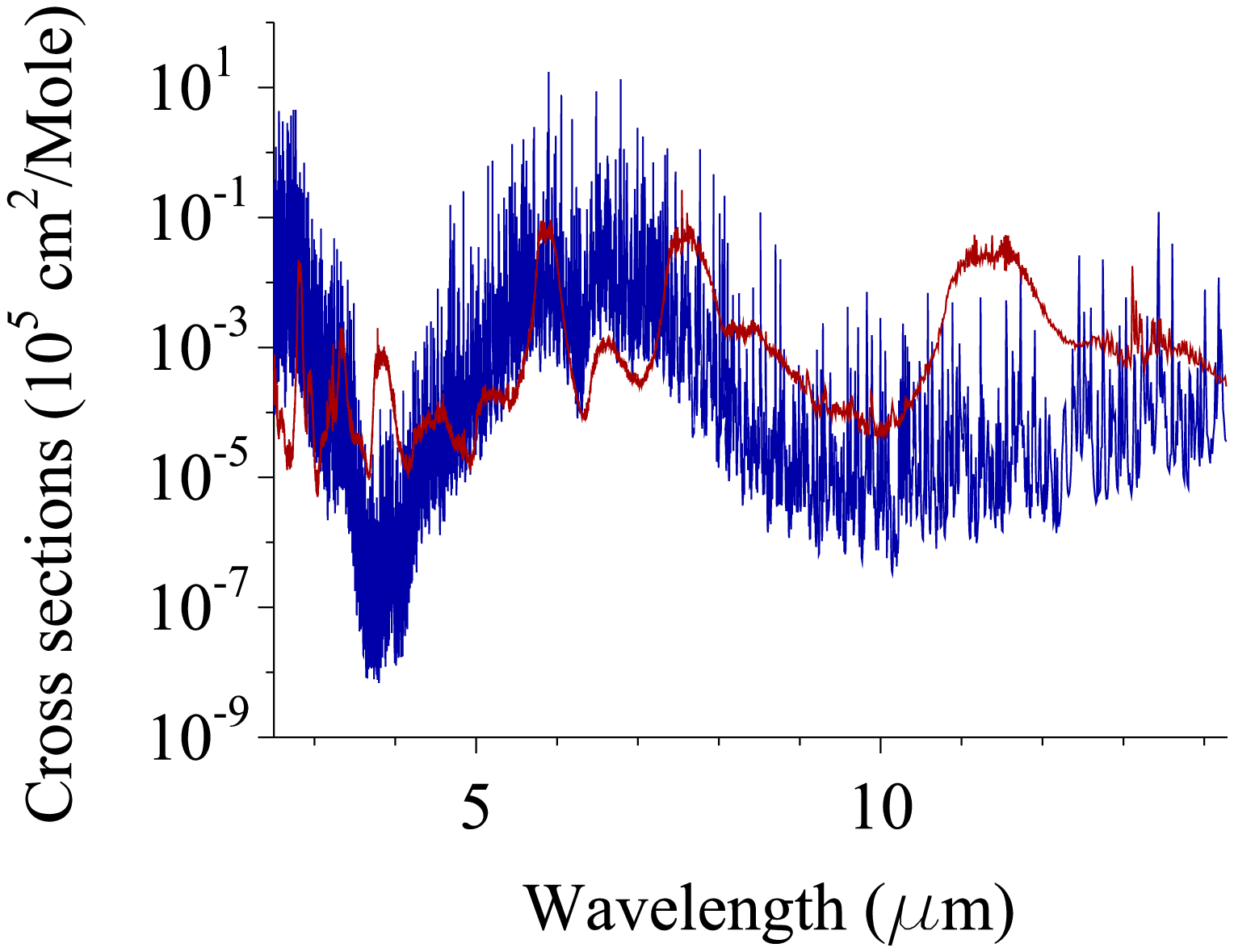}}
\caption{Temperature-dependent absorption spectra of a nitric acid (smooth curve) and
water vapour (spiky curve) mixture. Spectra
are for 2\%\ nitric acid at 300 K (upper) and 500 K (lower). The water spectra show a strong
window around 4 $\mu$m.}
\label{fig:spec2c}
\end{center}
\end{figure}

Figure~\ref{fig:PNNL} compares our calculations with the cross
sections given in the PNNL database \citep{PNNL}. More detailed
comparisons with both PNNL and HITRAN \citep{jt557} are given by
\citet{jt603}.

Given that HNO$_3$ is being considered as possible biomarker and that
any biomarker will probably have to be observed in a humid atmosphere,
we consider a number of mixed water -- nitric acid spectra.  Figure
\ref{fig:spec2c} gives an overview of such spectra which have been
constructed using water from HITRAN with 2~\%\ HNO$_3$ added for 300 K
and 500 K.  As is well-known, water absorptions dominate much of the
infrared. However, HNO$_3$ features are clearly visible at about 7.5
$\mu$m an d 11.2 $\mu$m. These are precisely where HNO$_3$ is clearly
seen in our own atmosphere from space \citep{96BlDe.HNO3}.

\begin{figure}
\begin{center}
{\leavevmode \epsfxsize=6.0cm \epsfbox{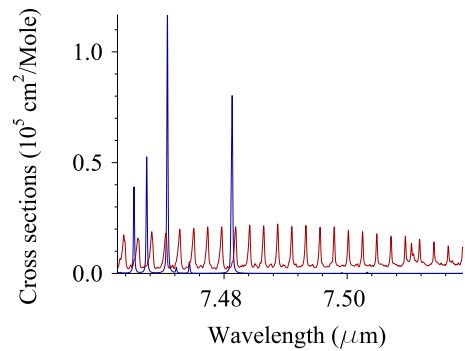}}
\caption{Absorption spectrum of a nitric acid (regular set of lines) and water vapour (four strong lines) mixture in the 1330 -- 1340 cm$^{-1}$ region. The spectrum
is for 300 K with a concentration of 2 \%\ nitric acid.}
\label{fig:spec7}
\end{center}
\end{figure}

\begin{figure}
\begin{center}
{\leavevmode \epsfxsize=6.0cm \epsfbox{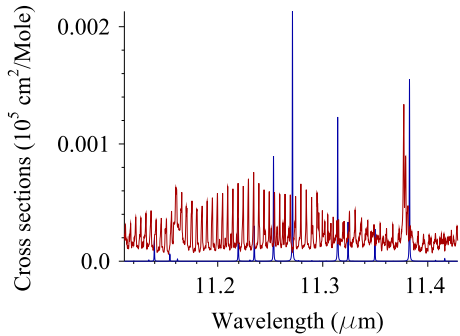}}
\caption{Absorption spectrum of a nitric acid (regular set of lines) and water vapour
(isolated lines) mixture in the 875 -- 900 cm$^{-1}$. The spectrum
is for 300 K with a trace  concentration of 0.01 \%\ nitric acid.}
\label{fig:spec11}
\end{center}
\end{figure}

Figures \ref{fig:spec7} and \ref{fig:spec11} give detailed comparisons
for the two regions in question. The spectrum in the 7.5 $\mu$m region
shows a clear, regular spectral signature of HNO$_3$ but would require
both a significant fraction of HNO$_3$ present and high resolution for
a successful detection. When viewed in the Earth's atmosphere this
region also shows strong absorption lines due to CO$_2$, although
there are not enough of these to mask the HNO$_3$. The 11.2 $\mu$m
region is actually much more promising. The figure is drawn for only a
trace concentration of HNO$_3$ (0.01 \%\ of water), much less than is
present in the Earth's atmosphere. Furthermore there is a distinct
HNO$_3$ band centred about 11.25 $\mu$m and a feature 11.38 $\mu$m
which should be visible (possibly blended together) at lower
resolution. This would appear to be the most promising region for an
HNO$_3$ detection. We note that there is one further HNO$_3$ band
clearly visible in Fig.~\ref{fig:spec2c} at about 3~$\mu$m. This
feature is a blend of the $2\nu_3$ overtone (band centre at 3404.4
\cm), and the $\nu_2 + \nu_4$ (2998.5 \cm) and $\nu_2 + \nu_3$ (3022.1
\cm) combibation bands. This compound band sits in a window in the
water spectrum but is not particularly prominent in the spectrum of
the Earth's atmosphere, presumably because it is actually
significantly weaker than the features at 11.2 $\mu$m and 7.5 $\mu$m.

\begin{table*}
\caption{ Extract from the state file for HNO$_3$. Full tables
are available from
http://cdsarc.u-strasbg.fr/cgi-bin/VizieR?-source=J/MNRAS/xxx/yy.}
\label{tab:levels}
\begin{center}
\footnotesize
\tabcolsep=5pt
\begin{tabular}{rrrrlccccccccc}
\hline
     $n$ & \multicolumn{1}{c}{$\tilde{E}$} &  $g$    & $J$  &  \multicolumn{1}{c}{$\Gamma$} &
$v_1$ & $v_2$ &$v_3$ &$v_4$ &$v_5$ &$v_6$ &$v_7$ &$v_8$ &$v_9$   \\
\hline
           1&     0.000000     & 6     & 0 &   A'       & 0 & 0 & 0 & 0 & 0 & 0 & 0 & 0 & 0\\
           2&   458.200000     & 6     & 0 &   A"       & 0 & 0 & 0 & 0 & 0 & 0 & 0 & 0 & 1\\
           3&   580.300000     & 6     & 0 &   A'       & 0 & 0 & 0 & 0 & 0 & 0 & 0 & 1 & 0\\
           4&   646.450000     & 6     & 0 &   A'       & 0 & 0 & 0 & 0 & 0 & 0 & 1 & 0 & 0\\
           5&   763.100000     & 6     & 0 &   A"       & 0 & 0 & 0 & 0 & 0 & 1 & 0 & 0 & 0\\
           6&   879.050000     & 6     & 0 &   A'       & 0 & 0 & 0 & 0 & 1 & 0 & 0 & 0 & 0\\
           7&   896.300000     & 6     & 0 &   A'       & 0 & 0 & 0 & 0 & 0 & 0 & 0 & 0 & 2\\
           8&  1038.000000     & 6     & 0 &   A"       & 0 & 0 & 0 & 0 & 0 & 0 & 0 & 1 & 1\\
           9&  1100.800000     & 6     & 0 &   A"       & 0 & 0 & 0 & 0 & 0 & 0 & 1 & 0 & 1\\
          10&  1148.975745     & 6     & 0 &   A'       & 0 & 0 & 0 & 0 & 0 & 0 & 0 & 2 & 0\\
          11&  1205.600000     & 6     & 0 &   A'       & 0 & 0 & 0 & 0 & 0 & 1 & 0 & 0 & 1\\
          12&  1213.857206     & 6     & 0 &   A'       & 0 & 0 & 0 & 0 & 0 & 0 & 1 & 1 & 0\\
          13&  1275.720926     & 6     & 0 &   A'       & 0 & 0 & 0 & 0 & 1 & 0 & 1 & 0 & 0\\
          14&  1289.000000     & 6     & 0 &   A"       & 0 & 0 & 0 & 0 & 0 & 0 & 0 & 0 & 3\\
          15&  1303.100000     & 6     & 0 &   A'       & 0 & 0 & 0 & 1 & 0 & 0 & 0 & 0 & 0\\
          16&  1325.650000     & 6     & 0 &   A'       & 0 & 0 & 1 & 0 & 0 & 0 & 0 & 0 & 0\\
          17&  1340.608559     & 6     & 0 &   A"       & 0 & 0 & 0 & 0 & 0 & 1 & 0 & 1 & 0\\
          18&  1343.600000     & 6     & 0 &   A"       & 0 & 0 & 0 & 0 & 1 & 0 & 0 & 0 & 1\\
          19&  1403.878302     & 6     & 0 &   A"       & 0 & 0 & 0 & 0 & 0 & 1 & 1 & 0 & 0\\
          20&  1450.282008     & 6     & 0 &   A'       & 0 & 0 & 0 & 0 & 1 & 0 & 0 & 1 & 0\\

\hline

\end{tabular}
\end{center}

\mbox{}\\
{\flushleft
$n$:   State counting number.     \\
$\tilde{E}$: State energy in \cm. \\
$g$: State degeneracy.            \\
$J$: Rotational quantum number.\\
$\Gamma$:   Total symmetry. \\
$v_1$, $v_2$, $v_3$, $v_4$, $v_5$, $v_6$, $v_7$, $v_8$, $v_9$:  vibrational quantum numbers.\\ }
\end{table*}

\begin{table}
\caption{ Extracts from the transitions file for HNO$_3$.
 Full tables are available from
http://cdsarc.u-strasbg.fr/cgi-bin/VizieR?-source=J/MNRAS/xxx/yy. }
\label{tab:trans}
\begin{center}
\begin{tabular}{ccc}
\hline\hline
$F$	&  $I$ 		& 		$A_{FI}$\\
\hline

    14516084  &   14516083  &   4.5130E--10.   \\
    14516265  &   14516264  &   8.1767E--10.   \\
    14515899  &   14515898  &   1.8398E--09.   \\
    14516082  &   14516081  &   2.4872E--08.   \\
    14516263  &   14516262  &   2.8596E--08.   \\
    14515897  &   14515896  &   7.7843E--08.   \\
    14516080  &   14516079  &   6.5506E--07.   \\
    14516261  &   14516260  &   6.2021E--07.   \\
    14515895  &   14515894  &   1.5810E--06.   \\
    14516078  &   14516077  &   7.7309E--06.   \\
    14516080  &   14516078  &   7.1042E--06.   \\
    14515895  &   14515893  &   1.5567E--05.   \\
    14516259  &   14516258  &   5.9759E--06.   \\
    14515897  &   14515895  &   1.8498E--05.   \\
    14515893  &   14515892  &   1.4427E--05.   \\
    14516261  &   14516259  &   1.2759E--05.   \\

\hline
\end{tabular}
\mbox{}\\
{\flushleft
$F$: Upper state counting number; \\
$I$:      Lower state counting number;\\
$A_{FI}$:  Einstein-A coefficient in s$^{-1}$.\\}

\end{center}
\end{table}


\section{Conclusions}

As part of the ExoMol project we have computed a comprehensive line
list for the possible biomarker nitric acid applicable to temperatures
up to 500 K. To do this we have had to develop a novel methodology
capable of treating the anharmonic motions of a heavy 5-atom molecule
\citep{jt588,jt603}. The line lists, which contains almost 7 billion
lines, can be downloaded from the CDS, via
ftp://cdsarc.u-strasbg.fr/pub/cats/J/MNRAS/, or
http://cdsarc.u-strasbg.fr/viz-bin/qcat?J/MNRAS/, or from
www.exomol.com.

Simulated spectra for a water-rich atmosphere suggest that there is real prospect
of detecting trace quantities of nitric acid in the 11.25 $\mu$m region.

\section*{Acknowledgements}

This work is supported by ERC Advanced Investigator Project 267219.
The authors acknowledge the use of the UCL Legion High Performance Computing
facility (Legion@UCL), and associated support services, in the completion of this work.

\bibliographystyle{mn2e}

\label{lastpage}

\end{document}